\documentclass[aps,physrev,twocolumn,superscriptaddress,nofootinbib,balancelastpage]{revtex4-2} 
\usepackage[latin1]{inputenc}
\usepackage{amsmath,amssymb}
\usepackage{mathrsfs} 
\usepackage{siunitx}
\usepackage{braket}
\usepackage{calc}
\usepackage{esint}
\usepackage{tabularx}
\usepackage{dsfont}
\usepackage{color}
\usepackage{ifthen}
\usepackage{graphicx}
\usepackage{seqsplit}
\usepackage[colorlinks,allcolors=black]{hyperref} 
\usepackage[capitalise]{cleveref}

\allowdisplaybreaks

\newcommand{\dif}{\mathrm{d}}%
\newcommand{\pdif}[2]{\frac{\partial#1}{\partial#2}}%
\newcommand{\Nabla}{\vec{\nabla}}%
\newcommand{\fdif}{\operatorname{\delta}}%
\newcommand{\Fdif}[2]{\frac{\fdif\!#1}{\fdif\!#2}}%

\newcommand{\rt}{(\vec{r},t)}

\newcommand{\ZT}[1]{\textquotedblleft#1\textquotedblright}%
\newcommand{\SmoluchowskiDistribution}{\widehat{\Psi}}

\newcolumntype{Y}{>{\centering\arraybackslash}X}%
\newcolumntype{Z}{>{\raggedright\arraybackslash}X}%

\newlength{\myl}%
\newcommand{\SUM}[2]{{\setlength{\myl}{\widthof{$\displaystyle\sum_{#1}^{#2}$}*\real{0.5}-\widthof{$\displaystyle\sum$}*\real{0.5}}\sum_{#1}^{#2}\;\hspace{-\the\myl}}}
\newcommand{\INT}[3]{\settowidth{\myl}{$\displaystyle\int_{#1}^{#2}$}{\int_{#1}^{#2}\;\;\;\hspace{-\the\myl}\dif #3}\,}
\newcommand{\TINT}[3]{\settowidth{\myl}{$\int_{#1}^{#2}$}{\int_{#1}^{#2}\!\ifthenelse{\equal{#1#2}{}}{}{\;\;\;\;\hspace{-\the\myl}}\dif #3}\,}%
\newcommand{\EINT}[3]{\settowidth{\myl}{$\int_{#1}^{#2}$}{\int_{#1}^{#2}\;\;\;\,\hspace{-\the\myl}\dif #3}\,}
\newcommand{\CINT}[3]{\settowidth{\myl}{$\displaystyle\int_{#1}^{#2}$}{\oint_{#1}^{#2}\;\;\;\hspace{-\the\myl}\dif #3}\,}

\begin{document}
\title{Perspective: New directions in dynamical density functional theory}

\author{Michael te Vrugt}
\affiliation{Institut f\"ur Theoretische Physik, Center for Soft Nanoscience, Westf\"alische Wilhelms-Universit\"at M\"unster, 48149 M\"unster, Germany}

\author{Raphael Wittkowski}
\email[Corresponding author: ]{raphael.wittkowski@uni-muenster.de}
\affiliation{Institut f\"ur Theoretische Physik, Center for Soft Nanoscience, Westf\"alische Wilhelms-Universit\"at M\"unster, 48149 M\"unster, Germany}

\begin{abstract}
Classical dynamical density functional theory (DDFT) has become one of the central modeling approaches in nonequilibrium soft matter physics. Recent years have seen the emergence of novel and interesting fields of application for DDFT. In particular, there has been a remarkable growth in the amount of work related to chemistry. Moreover, DDFT has stimulated research on other theories such as phase field crystal models and power functional theory. In this perspective, we summarize the latest developments in the field of DDFT and discuss a variety of possible directions for future research.
\end{abstract}
\maketitle

\section{Introduction}
Classical dynamical density functional theory (DDFT) is a theory for the time evolution of the one-body density $\rho$ of a fluid, which is based on extending results from equilibrium density functional theory (DFT) \cite{EbnerSS1976,SaamE1977} towards the nonequilibrium case. DDFT exists in deterministic and stochastic variants. Deterministic DDFT was first introduced phenomenologically by \citet{Evans1979} and later derived from microscopic particle dynamics by \citet{MarconiT1999}, \citet{ArcherE2004}, \citet{Yoshimori2005}, and \citet{EspanolL2009}. Stochastic DDFT was pioneered by \citet{Munakata1989}, \citet{Kawasaki1994}, and \citet{Dean1996}. Finally, Fraaije and coworkers \cite{Fraaije1993,FraaijevVMPEHAGW1997} have developed a DDFT for polymers. Nowadays, DDFT has been applied in a large and diverse number of fields ranging from simple \cite{Archer2005} and colloidal \cite{MarconiT1999} fluids to plasmas \cite{DiawM2016} and microswimmers \cite{MenzelSHL2016}. A recent review can be found in Ref.\ \cite{teVrugtLW2020}.

Apart from its \ZT{older brother} DFT, DDFT has two younger \ZT{siblings} -- namely phase field crystal (PFC) models \cite{ElderKHG2002,EmmerichEtAl2012} and power functional theory (PFT) \cite{SchmidtB2013,Schmidt2022}. PFC models are simpler than DDFT (and can be derived from it \cite{vanTeeffelenBVL2009}), whereas PFT is more complex and contains DDFT as a limiting case. The development of these two other theories is intimately connected to that of DDFT, and progress in DDFT has stimulated progress in these other theories. For example, the development of active DDFT \cite{WensinkL2008} allowed to derive an active PFC model \cite{MenzelL2013,MenzelOL2014}, and improvements in DDFT allow also for more accurate PFT equations \cite{WittmannLB2020}.

This perspective article complements our review \cite{teVrugtLW2020} in two ways. First, we present articles on DDFT from the past two years, thereby also covering articles not included in the review because they are too recent. Second, we discuss perspectives for future work, thereby providing also a more speculative outlook on the new directions the field is developing towards.

This article is structured as follows: In \cref{sec:derivation}, we briefly review the derivation of DDFT. Recent developments are explained in \cref{recent}. In \cref{perspective}, we discuss possible future directions. We conclude in \cref{conc}.

\section{\label{sec:derivation}A brief introduction to DDFT}
DDFT is an extension of classical \textit{density functional theory} (DFT), which describes the equilibrium state of a classical fluid. Classical DFT, in turn, originates from the more widely known quantum DFT developed by \citet{HohenbergK1964}, which allows to model the ground state of a many-electron system.

We start by briefly introducing DFT following Refs.\ \cite{Loewen1994a,teVrugtLW2020}. The microscopic description of a classical many-body system requires, in principle, knowledge of the exact phase-space distribution function. Classical DFT makes use of the fact that the state of an equilibrium fluid is completely determined once the one-body density $\rho$ is known. The equilibrium density $\rho_{\mathrm{eq}}$ can be calculated from the grand-canonical free energy functional $\Omega$ (depending on the temperature $T$ and the chemical potential $\mu$) via the minimization principle (\textit{DFT equation})
\begin{equation}
\frac{\delta \Omega(T,\mu,[\rho])}{\delta \rho(\vec{r})}\bigg|_{\rho = \rho_{\mathrm{eq}}} = 0.
\label{dft}
\end{equation}
From the grand-canonical functional $\Omega$, the canonical free energy functional $F$ can be obtained via the Legendre transformation
\begin{equation}
\Omega(T,\mu,[\rho]) = F(T,[\rho]) - \mu\INT{}{}{^3r}\rho(\vec{r}).
\end{equation}
The free energy $F$ can be split into three parts:
\begin{equation}
F(T,[\rho])= F_{\mathrm{id}}(T,[\rho]) + F_{\mathrm{exc}}(T,[\rho]) + F_{\mathrm{ext}}([\rho]).
\label{freeenergy}
\end{equation}
The first term in \cref{freeenergy} is the exactly known ideal gas free energy
\begin{equation}
F_{\mathrm{id}}(T,[\rho]) = k_B T \INT{}{}{^3r}\rho(\vec{r})(\ln(\Lambda^3\rho(\vec{r}))-1).
\label{idealfreeenergy}
\end{equation}
Here, $k_B$ is the Boltzmann constant and $\Lambda$ is the (irrelevant) thermal de Broglie wavelength. The third term in \cref{freeenergy} is the external free energy
\begin{equation}
F_{\mathrm{ext}}([\rho]) = \INT{}{}{^3r}\rho(\vec{r})U_1(\vec{r})
\end{equation}
depending on the external potential $U_1$. Finally, the excess free energy $F_{\mathrm{exc}}$ describes interactions of the particles in the system and is not known exactly. (Parametric dependencies are suppressed from here on.)

Now, we turn to the nonequilibrium case and present the derivation of DDFT following \citet{ArcherE2004}. The starting point is the Smoluchowski equation
\begin{equation}
\begin{split}
&\pdif{}{t}\SmoluchowskiDistribution(\{\vec{r}_k\},t)\\
&=\Gamma\sum_{i=1}^{N}\vec{\nabla}_{\vec{r}_i}\cdot(k_B T \vec{\nabla}_{\vec{r}_i} + \vec{\nabla}_{\vec{r}_i} U(\{\vec{r}_k\},t))\SmoluchowskiDistribution(\{\vec{r}_k\},t)
\end{split}
\label{smoluchowski}
\end{equation}
describing the dynamics of the distribution function $\SmoluchowskiDistribution$ depending on the positions $\vec{r}_k$ of the $N$ particles (we consider spherical overdamped particles with two-body interactions only) and on time $t$. Here, $\Gamma$ is the mobility of a particle and $U=U_1+U_2$ (with the pair-interaction potential $U_2$) the total potential. The one-body density is defined as
\begin{equation}
\rho(\vec{r}_1,t)=N\INT{}{}{^3 r_2}\dotsb \INT{}{}{^3 r_N}\SmoluchowskiDistribution(\{\vec{r}_k\},t).
\label{obdensity}
\end{equation}
Integrating \cref{smoluchowski} over the coordinates of all particles except for one and using \cref{obdensity} gives
\begin{equation}
\begin{split}
\pdif{}{t}\rho\rt &= D \Nabla^2 \rho\rt + \Gamma\Nabla\cdot(\rho\rt\Nabla U_1\rt)\\
&\quad\:\!+ \Gamma \vec{\nabla}\cdot\INT{}{}{^3r'}\rho^{(2)}(\vec{r},\vec{r}',t)\vec{\nabla}U_2(\vec{r},\vec{r}')
\end{split}
\label{deterministicnbody}
\end{equation}
with the diffusion constant $D=\Gamma k_B T$, where we write $\vec{r}$ for $\vec{r}_1$ and $\vec{r}'$ for $\vec{r}_2$. Since \cref{deterministicnbody} depends also on the unknown two-body density $\rho^{(2)}$, we require a closure. For this purpose, one uses the \textit{adiabatic approximation}, which corresponds to the assumption that the correlations in the system are the same as in an equilibrium system. This allows to insert the equilibrium relation
\begin{equation}
\rho(\vec{r})\vec{\nabla}\frac{\delta F_{\mathrm{exc}}[\rho]}{\delta \rho(\vec{r})}=\INT{}{}{^3r'}\rho^{(2)}(\vec{r},\vec{r}')\vec{\nabla}U_2(\vec{r},\vec{r}')
\end{equation}
into \cref{deterministicnbody} to obtain the \textit{DDFT equation}
\begin{equation}
\pdif{}{t}\rho\rt = \Gamma\vec{\nabla}\cdot\bigg(\rho\rt\vec{\nabla}\frac{\delta F[\rho]}{\delta \rho\rt}\bigg).
\label{ddfteq}
\end{equation}
Important alternative derivation routes start from the Langevin equations \cite{MarconiT1999,MarconiT2000} that describe the motion of the particles in the system or use the Mori-Zwanzig formalism \cite{Yoshimori2005,EspanolL2009}. A complete overview is given in Ref.\ \cite{teVrugtLW2020}.

\textit{Phase field crystal (PFC) models} \cite{ElderKHG2002,EmmerichEtAl2012} are a closely related approach. They are based on an order parameter $\psi$ that is related to the density $\rho$ by $\rho=\rho_0(1+\psi)$, where $\rho_0$ is a spatially and temporally constant reference density. The governing equation of PFC models is given by
\begin{equation}
\pdif{}{t} \psi\rt = M\Nabla^2\Fdif{F[\psi]}{\psi\rt}    
\end{equation}
(with a mobility $M$) and can be derived from \cref{ddfteq} by making the approximation of a constant mobility. The free energy $F$ in PFC models is also considerably simpler than that of (D)DFT and can be derived by performing a Taylor expansion for the logarithm in \cref{idealfreeenergy} and a functional Taylor expansion combined with a gradient expansion for the excess free energy $F_{\mathrm{exc}}$. A detailed discussion of this derivation can be found in Refs.\ \cite{teVrugtLW2020,EmmerichEtAl2012,ArcherRRS2019}.

An extension of DDFT that has gained some popularity is \textit{power functional theory} (PFT), which was developed by \citet{SchmidtB2013} (see Refs.\ \cite{Schmidt2022,teVrugtLW2020,Schilling2021} for a review). PFT describes the nonequilibrium dynamics of many-body systems and is, like DFT, a formally exact variational theory. The variational principle is formulated here not for the density $\rho$, but for the current $\vec{J}$ that minimizes the so-called \ZT{power functional}. This functional can be split into an \ZT{ideal part} (the part that is already present in DDFT) and an \ZT{excess part} $P_{\mathrm{exc}}$. This leads to the governing equation \cite{SchmidtB2013}
\begin{equation}
\frac{\vec{J}\rt}{\Gamma \rho\rt} + \frac{\delta P_{\mathrm{exc}}([\rho,\vec{J}],t)}{\delta \vec{J}\rt} = - \vec{\nabla}\frac{\delta F[\rho]}{\delta\rho\rt}
\label{fundamentaleqpft}
\end{equation}
of PFT, which reduces to \cref{ddfteq} for $P_{\mathrm{exc}}=0$.

\begin{figure}
\centering
\includegraphics[width=\linewidth]{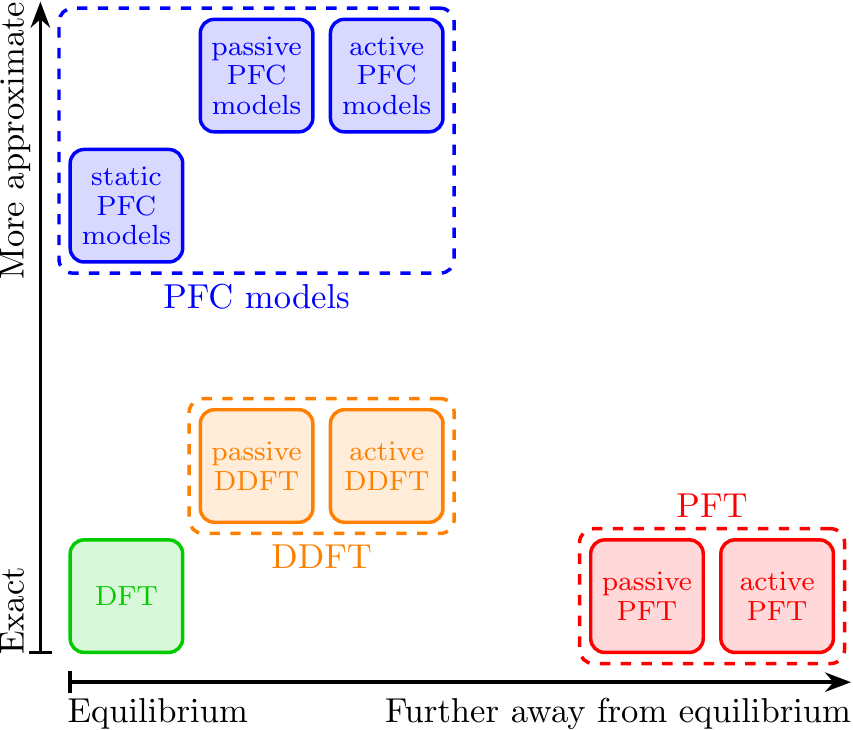}
\caption{\label{fig:schaubild}Relations of the various theories discussed in \cref{sec:derivation}. Different approaches can be used depending on the desired accuracy and the importance of nonequilibrium effects. The top right corner leaves room for future developments.}
\end{figure}

The relations between DFT, DDFT, PFC models, and PFT are visualized in \cref{fig:schaubild}. DFT is an exact theory (apart from approximations required for the free energy functional) for an equilibrium fluid. PFC models also allow to describe equilibrium systems, but with a more approximate free energy functional. If we go to the nonequilibrium case and use DDFT or PFC models, we are making an approximation (namely the adiabatic approximation), such that the theory is not exact. PFT, finally, provides an exact nonequilibrium theory. For the three dynamical theories DDFT, PFC models, and PFT, we include also an active variant which is based on the same sorts of approximations, but will usually be applied to systems further away from equilibrium (namely active ones \cite{MarchettiJRLPRS2013,BechingerdLLRVV2016}).

\section{\label{recent}Recent developments in DDFT}

Since its original development, DDFT has found a very remarkable number of applications. A detailed overview has been given in our recent review article \cite{teVrugtLW2020}. Since the purpose of the present manuscript is to highlight more recent developments and, in particular, future perspectives, we now briefly discuss the work on DDFT from the past two years, thereby covering (though not exclusively) articles not yet discussed in our review. 

\begin{figure*}
\centering
\includegraphics[width=\linewidth]{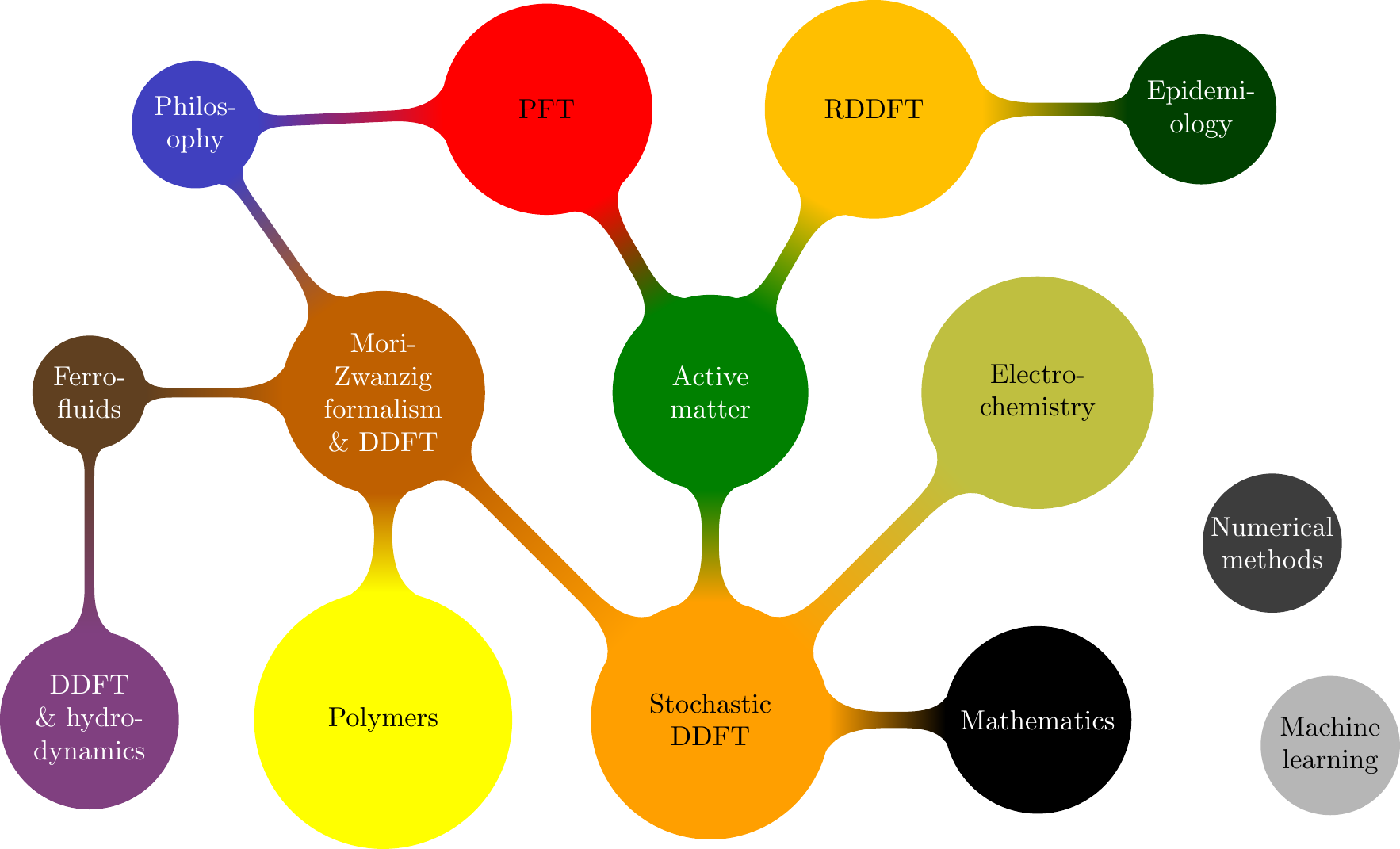}
\caption{\label{fig:kreise}DDFT-related articles published on selected topics in 2020-2022. The area of the circles indicates the number of articles, connections show that some articles belong to several topics. Related colors indicate related topics.}
\end{figure*}

The amount of articles published on certain selected topics is visualized in \cref{fig:kreise}. While for some topics the number of existing articles mainly originates from the work of a single author or research group, it is possible to identify certain trends. Perhaps the most interesting one is chemistry, which is addressed by a variety of authors in a variety of ways.

\subsection{Epidemiology}
Already in 2020, the worldwide outbreak of the coronavirus disease COVID-19 has motivated the application of DDFT to disease spreading. In the \textit{SIR-DDFT} model (a combination of the susceptible-infected-recovered (SIR) model \cite{KermackM1927} with DDFT), repulsive particle interactions are used to represent social distancing measures \cite{teVrugtBW2020}. The SIR-DDFT model has been extended to model governmental intervention strategies that can lead to multiple waves of a pandemic \cite{teVrugtBW2020b}. This extension represents the first DDFT with a time-dependent interaction potential. Moreover, a software package has been developed to simulate epidemic outbreaks in the SIR-DDFT model \cite{JeggleW2021}. \citet{YiXJ2021} have proposed a way of combining the SIR-DDFT model with WiFi data in order to get estimates for values of the model's parameters. Some further extensions were suggested in Ref.\ \cite{DuranK2021}. A brief overview was given in Ref.\ \cite{Loewen2021b}.

\subsection{Chemistry}
The SIR-DDFT model has the mathematical structure of a reaction-diffusion DDFT (RDDFT) \cite{Lutsko2016,LutskoN2016}, i.e., a reaction-diffusion equation \cite{Turing1952} with the diffusion terms replaced by the right-hand side of \cref{ddfteq}. RDDFT has recently been used also to study actively switching Brownian particles \cite{MonchoD2020,BleyDM2021,BleyHDM2022}. Here, it is assumed that the particles can switch between two states with different sizes at a certain rate. This work also explicitly compares RDDFT to Brownian dynamics simulations and reports good agreement \cite{BleyDM2021}. Another active matter model based on RDDFT has been developed by \citet{AlstonPVB2022}. Finally, RDDFT has also been applied in actual chemistry to study catalytic oxidation \cite{LiuL2020}, crystal nucleation \cite{ZongWHWLH2022}, metal corrosion \cite{ChenLL2022}, reactions on catalytic substrates \cite{TangYZQXZ2021}, and reactions on electrode surfaces \cite{Liu2020}.

Also apart from RDDFT, quite a number of recent applications of DDFT come from chemistry and chemical physics, broadly construed. In particular, DDFT has been used in electrochemistry  to model systems and processes such as charging of electric double layers \cite{QingZW2021,MaJLvR2022} and supercapacitors \cite{QingJ2022,AslyamovSA2022}, counterions \cite{Frusawa2020}, dielectricity \cite{QingLZTQMXZ2020,Dejardin2022}, electrolytes \cite{MahdisoltaniG2021,AslyamovJ2022,Frusawa2022,Frusawa2022b}, impedance response \cite{TomlinRKMG2021}, and ion adsorption \cite{ZhaoQLXZL2021}. A further example is solvation dynamics \cite{LiZQYXLZ2021,LiQYPXLZ2021}, which has a long tradition as an application of DDFT \cite{ChandraB1991}. Finally, DDFT can be used to study nanoparticle separation \cite{YuWLLTBZ2021}, the release of molecules from nanoparticles \cite{MonchoEtAl2020} or porous surfaces \cite{ChangY2020}, and wound healing \cite{LiWLHW2020}.

DDFT for polymer systems \cite{Fraaije1993,FraaijevVMPEHAGW1997}, which is already well established in chemical physics, should of course also be mentioned here. On the theoretical side, the microscopic construction of mobility functions was studied \cite{LiDS2021,ManthaQS2020,SchmidL2020}. Further work considered the influence of correlations on polymer dynamics \cite{ChenQZY2020}, memory effects \cite{RottlerM2020,Muller2022}, micelle relaxation \cite{PantelidouGFAM2022}, and morphological phase transitions \cite{MartinezAugustinaRSZMRM2022}. The relation to other relaxation models was briefly discussed in Ref.\ \cite{ErukhimovichKK2022}. Finally, the MesoDyn software \cite{AltevogtEFMvV1999}, which allows to simulate polymer systems based on DDFT, remains an important tool in the study of polymer dynamics \cite{PangJDZDGNQQL2021,ChaeLBSPKL2021,SlimaneGGC2020,LeePCLLYKJK2021,FengW2022}.

\subsection{Theoretical developments}
Classical DFT is formulated in the grand-canonical ensemble, which is inappropriate for very small closed systems and somewhat inconsistent with the fact that DDFT has the form of a (particle-conserving) continuity equation. Work to address this issue has been directed at formulating a canonical DFT \cite{Lutsko2022,delasHerasS2014,WhiteV2001,WhiteG2002} and at extending DDFT towards the canonical case in a formalism known as \ZT{particle-conserving dynamics} (PCD) \cite{delasHerasBFS2016}. \citet{SchindlerWB2019}, who developed a PCD for mixtures, have noted that this theory makes the unphysical prediction of allowing hard rods in one dimension to pass through each other. This problem is solved in \ZT{order-preserving dynamics} (OPD) \cite{WittmannLB2020}, a variant of PCD based on an asymmetric interaction potential.

OPD has also been of interest for philosophers of physics. Since it treats observationally indistinguishable particle configurations in different ways, it is of relevance for the long-standing philosophical debate concerned with whether such distinctions are possible \cite{teVrugt2021b}. More generally, DDFT has been discussed in philosophy in relation to the problem of thermodynamic arrow of time, i.e., the question how the irreversibility of macroscopic thermodynamics is compatible with the reversibility of the microscopic laws of physics \cite{teVrugt2020,teVrugt2021,teVrugtTW2021}, and to analyze the problem of scientific reduction \cite{teVrugtNS2022}. In particular, Ref.\ \cite{teVrugt2020} had the specific aim of developing a philosophy of DDFT.

Some of these discussions \cite{teVrugt2020,teVrugt2021} discuss DDFT in relation to the \textit{Mori-Zwanzig formalism} \cite{Mori1965,Nakajima1958,Zwanzig1960,Grabert1982,teVrugtW2019d}, which allows to derive irreversible transport equations from reversible microdynamics and thus to understand irreversibility \cite{teVrugt2021,Zeh1989}. This formalism has also been used to derive \cite{Yoshimori2005,EspanolL2009} and extend \cite{WittkowskiLB2012,WittkowskiLB2013,AneroET2013,CamargodlTDZEDBC2018} DDFT. It is also important for more recent work. In particular, it plays a prominent role in Fang's development of a DDFT for ferrofluids \cite{Fang2019,Fang2020,Fang2022,Fang2022b}, in analyzing memory effects in polymers \cite{RottlerM2020,Muller2022}, and in the study of crystal elasticity \cite{RasSF2020,Haussmann2022,GangulySLHHKOF2022}. A stochastic theory related to stochastic DDFT was derived using the Mori-Zwanzig formalism in Ref.\ \cite{Uneyama2022}. Since the Mori-Zwanzig formalism continues to be improved \cite{GlatzelS2021,MeyerVS2019,teVrugtW2019,teVrugtHW2021}, it is likely to play an important role also in future work on DDFT.

\subsection{Applications in new physical contexts}
Simple and colloidal fluids remain a central field of application for DDFT, although more recent work on these systems has gone beyond \ZT{standard} DDFT in several ways. For example, \citet{MaroltR2020} have used DDFT to study colloids with Casimir and magnetic interactions. \citet{JiaK2021} used an extended form of DDFT \cite{AneroET2013} to model nonisothermal hard spheres. The transport of soft Brownian particles was analyzed by \citet{AntonovRM2021}. \citet{MontanezQG2021} studied diffusion on spherical surfaces. The nonequilibrium self-consistent generalized Langevin equation \cite{RamirezGonzalezM2010}, an extension of DDFT, was applied to arrested density fluctuations by \citet{LiraVR2021,LiraEscobedoVR2022}. Density fluctuations were also modeled using an adiabatic approximation by \citet{Szamel2022}. Finally, \citet{SharmaNB2022} have studied the local softness parameter (which is useful for the description of caging) using DDFT.

In addition, stochastic DDFT \cite{Dean1996,Kawasaki1994}, commonly referred to as \ZT{Dean-Kawasaki equation} (a name that is somewhat unfortunate as it fails to acknowledge the differences between Dean's and Kawasaki's approaches \cite{teVrugtLW2020}), has remained an important tool in the study of interacting particles with stochastic dynamics. Recent examples include active matter \cite{MartinOCFNTvW2021,ZakineFvW2020}, chemotaxis \cite{MahdisoltaniZDGG2021,BenAliZinatiDMGG2022}, electrolytes \cite{MahdisoltaniG2021,Frusawa2022,Frusawa2022b}, densely packed spheres \cite{Frusawa2021}, and proteins \cite{GoutalandvWFN2021}. \citet{Satin2022} suggested a link between stochastic DDFT and theories of gravity.

Moreover, there have been several extensions of DDFT towards physical systems not previously considered in the context of DDFT. Building up on earlier work \cite{Fang2019,Fang2020}, Fang \cite{Fang2022,Fang2022b,Fang2022c} has recently derived a DDFT for ferrofluids. Another example is the development of a DDFT for granular media \cite{Hurst2020,GoddardHO2021}. \citet{StantonOCISLG2021} have modeled cellular membranes in DDFT. These can be described as a mixture of lipids and proteins. Finally, \citet{WittmannSL2022} have derived a DDFT that allows to describe mechano-sensing in growing bacteria colonies.

DDFT is also frequently studied in relation to hydrodynamics. Several works have investigated the relation between DDFT and the Navier-Stokes equation \cite{QiaoZYQBSL2021,Mills2020,ZhaoQXZ2021}. \citet{StierleG2021} have derived a \ZT{hydrodynamic DFT}, which describes underdamped mixtures. An inertial DDFT with hydrodynamic interactions was studied by \citet{GoddardMS020}. Moreover, DDFT allows to model drying colloidal films \cite{HeMRTA202}, droplets \cite{PerezRGT2021}, flow in nanopores \cite{ZhaoQLXZL2021}, hydrodynamics of ferrofluids \cite{Fang2022c}, and polymer mixtures \cite{HowardN2020}.

\subsection{Mathematics and software}
DDFT is of interest not only in physics and chemistry, but also for applied mathematics and software development. Recent work has considered stochastic DDFT (the Dean-Kawasaki equation) \cite{CornalbaSZ2020,CornalbaSZ2021,LeDoussal2022,LeeY2020,HelfmannDDWS2021,CornalbaF2021,FehrmanG2021} and the McKean-Vlasov equation (a DDFT-type model) \cite{GomesPV2020,BechtoldC2021,DelgadinoGP2021,ZakineVE2022} from a mathematical perspective. Moreover, numerical methods were developed for DDFT \cite{CarrilloCKP2021,KrukCK2021,MendesRPK2021,BavnasGV2020,AduamoahGPR2020,GoddardGPS2020} and PFC models \cite{WangHW2021,WangH2021b,CoelhoVPM2021,LiM2021}. A particularly rapidly growing subfield is the application of machine learning, which can be used to learn static free energy functionals \cite{LinO2019,LinMO2020,CatsKdWvDCDvR2021,YatsyshinKD2022} that can be used also in DDFT. However, machine learning is also used in the dynamical case \cite{AhnEtAl2020,BhatiaEtAl2021,IngolfssonEtAl2022,ZhaoBB2021,ZhaoSBB2020,BhatiaEtAl2021b}. An example of the latter type is multiscale modeling of proteins based on machine learning in Ref.\ \cite{IngolfssonEtAl2022}, where the DDFT from Ref.\ \cite{StantonOCISLG2021} is used as a macroscale model.

\subsection{Related theories: PFC and PFT}
Recent studies of PFC models have focused on active matter with \cite{AroldS2020b,AroldS2020,teVrugtJW2021} and without \cite{OphausKGT2020,OphausKGT2021,HuangML2020,KrauseV2021,HollAGKOT2020,teVrugtHKWT2022} inertia, bifurcation diagrams \cite{HollAGKOT2020,teVrugtHKWT2022,HollAT2020,Knobloch2020}, colored noise \cite{AnkudinovSKYYG2021}, cubic terms \cite{ChenHHXHR2021}, crystals \cite{BackofenSWV2021}, dislocation lines \cite{SkogvollASSV2021}, electromigration \cite{WangGP2021}, grain boundaries \cite{BlixtH2021,MartineLaBoissoniereCL2022}, mixtures \cite{teVrugtHKWT2022,AnkudinovG2022,ShuaiWMTKD2021,LiM2021,SalvalaglioVHE2021}, nucleation \cite{PodmaniczkyG2021}, solidification \cite{AnkudinovEG2020,WangW2021}, and stress tensors \cite{SkogvollSA2021}.

The past two years have seen a significant amount of work on PFT and superadiabatic forces (forces that are not captured within the adiabatic approximation). On the one hand, the formalism has found several applications in the study of acceleration viscosities \cite{RennerSdlH2022}, active matter \cite{HermanndlHS2021}, the dynamics of the van Hove function \cite{TreffenstadtS2021,TreffenstadtSS2022}, shear flow \cite{JahreisS2020}, and superadiabatic demixing \cite{GeigenfeinddlHS2020}. On the other hand, there have been more theoretical developments such as the derivation of Noether's theorem for statistical mechanics \cite{HermannS2021,HermannS2021b,HermannS2022} (which also served as the basis for a \ZT{force-based DFT} \cite{TschoppSHMB2022}), a classification of nonequilibrium forces \cite{delasHerasS2020}, a custom flow method \cite{RennerSdlH2021}, philosophical investigations of PFT \cite{teVrugt2020,teVrugt2021b}, and a reassessment of the original derivation of PFT \cite{LutskoO2021}.

\subsection{Overview articles}
Finally, further overview articles covering (also) DDFT have been published; in particular an extensive review of PFT by \citet{Schmidt2022}, a tutorial on active DDFT by \citet{Loewen2021}, and several reviews on biology and medicine \cite{ReinhardtEtAl2022,Casalini2021,GiniunaiteBKM2020,AchaziHBDKMM2021,CostaV2021,BarbeeWATPK2021}, coarse-graining \cite{KlipensteinTJSvdV2021,Schilling2021}, electrochemistry \cite{SuYZ2021,JiangZCHL2020,TaoLL2020,JeanmairetRS2022}, multiscale modeling \cite{BiEtAl2021}, PFC models \cite{StarodumovAN2022}, and polymers \cite{GinzburgH2021,ChoiP2020,UthaleDSK2021,Muller2020,OkVS2021,SharmaD2022} in which DDFT is mentioned. In our view, this large number of overview articles published in two years further highlights how timely the topic is.

\section{\label{perspective}Perspectives for the future}

\subsection{Phase field crystal models}
The relation of PFC models to DDFT is a very complex one whose understanding demands further work. Actually, it is not even clear how to draw the boundary between them. Some authors see PFC models simply as a special case of DDFT \cite{TothTPTG2010}, some see the difference in the fact that PFC models use a gradient expansion for the excess free energy \cite{vanTeeffelenBVL2009}, while others reserve the name \ZT{PFC} for models that also have a constant mobility approximation and an expanded logarithm in the ideal gas free energy \cite{ArcherRRS2019}.

DFT functionals can, at least for hard particles, be derived pretty much \ZT{ab initio}. Fundamental measure theory (FMT) \cite{Roth2010} provides highly accurate expressions for the free energy functional in hard-particle systems, such that, if we know the particle shapes (and other basic parameter such as the temperature), we can construct the DDFT equation \eqref{ddfteq} without having to adjust any free parameters. In contrast, the free energy in PFC models is typically just assumed to have a very simple Swift-Hohenberg-type form \cite{SwiftH1977,EmmerichEtAl2012}, and the parameters of this free energy can then be adjusted to fit a wide class of materials \cite{ArcherRRS2019}. Nevertheless, a derivation of PFC models from DDFT does give microscopic expressions for all these parameters, and so \textit{in principle}, assuming the free energy to be known for a certain interaction, PFC models also do not contain any free parameters. However, this option is almost never used in practice. This has to do with the fact that the predictions of DFT for the PFC parameters can turn out to be quite inaccurate as a consequence of the fact that the approximations made in the derivation of PFC models from DDFT ($\psi$ is assumed to be small and slowly varying in space) are not well justified \cite{JaatinenAEA2009}.

Thus, more work is required in understanding the microscopic origins of PFC models from DDFT. This might allow for more accurate predictions of model parameters, the development of more accurate PFC models (and perhaps also phase field models \cite{Emmerich2008,SteinbachPNSPSR1996}, which are also connected to DDFT \cite{MauriB2021}), and in general a better understanding of scientific reduction \cite{teVrugtNS2022}. Recently, some work has been done in this direction. This includes a microscopic extension of the active PFC model towards mixtures \cite{teVrugtHKWT2022}, the development of a framework for obtaining gradient-based free energies from more general expressions \cite{LiMRBHS2022}, and in particular a systematic assessment of the derivation of PFC models from DDFT by \citet{ArcherRRS2019}, who argued that the order parameter $\psi$ of PFC models should be interpreted not as the dimensionless deviation of the density from a reference value, but as the logarithm of the density.

\subsection{Power functional theory}
Since, as explained in \cref{sec:derivation}, PFT contains all of DDFT, but also adds additional structure, it can be quite complex. If one is interested in a model that allows to describe far-from-equilibrium processes but that is also easy to handle, one could also use PFC approaches to approximate the DDFT terms in \cref{fundamentaleqpft}. This would allow to obtain a model that combines the simplicity of the PFC approach with the ability of PFT to model far-from-equilibrium processes, and would allow, e.g., to study memory in active matter within the PFC framework (as done phenomenologically in Ref.\ \cite{teVrugtJW2021}). Such a theory would fit in the currently empty spot at the top right of \cref{fig:schaubild}. A further interesting idea, suggested in Ref.\ \cite{BaulGBD2021}, would be to combine PFT with RDDFT (see \cref{recent}) in order to model far-from-equilibrium effects in chemical reactions.

On the other hand, also the theoretical foundations of PFT merit further investigation. In particular, a recent article by \citet{LutskoO2021} has highlighted certain issues in the original derivation of PFT by \citet{SchmidtB2013}. More generally, the usefulness of PFT in practice strongly depends on the availability of a good approximation for the excess power functional. Something that would significantly increase the power of PFT would be the development of something like an FMT for the excess power functional, which provides an accurate expression obtained from first principles. Moreover, as discussed in Refs.\ \cite{WittmannLB2020,teVrugt2021b}, the question whether a particular effect is to be classified as superadiabatic or not can strongly depend on the choice of the underlying equilibrium framework (e.g., on whether or not one uses OPD in one dimension), since this framework affects the effects of the adiabatic approximation.

\subsection{Active matter}
\textit{Active matter physics} \cite{MarchettiJRLPRS2013,BechingerdLLRVV2016}, the study of systems that contain self-propelled particles, continues to be a rapidly growing subfield of soft matter physics in which a number of interesting effects are presumably still to be discovered. Active particles can be described in DDFT using a one-body density that depends also on the orientation of the particles. Apart from this, the general idea behind the derivation (see \cref{sec:derivation}) is still the same. Active DDFT has a number of interesting applications, in particular in the study of microswimmers \cite{MenzelSHL2016,HoellLM2017,HoellLM2018,HoellLM2019} (see Ref.\ \cite{Loewen2021} for an overview). Moreover, active DDFT serves as the basis for the derivation of active PFC models \cite{MenzelL2013,MenzelOL2014,teVrugtHKWT2022}.

A conceptual challenge in modeling active particles using DDFT is that, as explained in \cref{sec:derivation}, DDFT is based on the assumption that the two-body correlations are the same as in an equilibrium system. Therefore, DDFT is based on a close-to-equilibrium assumption, which is problematic since active systems are far from equilibrium. This problem is, as mentioned in Refs.\ \cite{BickmannW2019b,teVrugtFHTW}, inherited by active PFC models. \citet{DhontPGB2021} have argued that active DDFT is inappropriate for particles with steep and short-ranged interactions. Moreover, DDFT models for microswimmers \cite{MenzelSHL2016,HoellLM2017} can become inaccurate at higher densities \cite{BickmannBW2022} since hydrodynamic interactions are modeled using a far-field approximation. PFT allows, in principle, to overcome the low-activity limitation as it does not require a close-to-equilibrium assumption, although the governing equation \eqref{fundamentaleqpft} of PFT in practice typically takes the form \ZT{DDFT equation + correction term}. Consequently, PFT has been successfully applied to active phase separation \cite{HermanndlHS2021}. Microscopically derived active matter models generally require as an input knowledge (or assumptions) about the correlations in the system \cite{JeggleSW2020}, and it is among the main virtues of DDFT that it provides such an input. Therefore, a promising direction would be to develop a DDFT-like theory based on correlations from a nonequilibrium steady state. Ideas of this form have been used in Refs.\ \cite{WittmannB2016,PototskyS2012,FarageKB2015}.

From a more \ZT{applied} perspective, an interesting project could be the study of topological defects in active matter using DDFT. For equilibrium systems, it has been found that DFT provides a quantitatively accurate description (as compared to experiments) of the topology of confined smectics \cite{WittmannCLA2021}. Given that topological defects are of central importance for the understanding of active matter systems \cite{ShankarSBMV2020}, this suggests the investigation of defect dynamics in active matter systems as a further application of active DDFT. Since even the topology of equilibrium smectics remains a topic of active research \cite{MonderkampWCASL2021,MonderkampWtVVWL2022}, the nonequilibrium case (that can be accessed by DDFT) promises even more interesting discoveries. A first step in this direction is the application of an active PFC model to this problem \cite{HuangLV2022}.

\subsection{Biology}
Closely related to active matter are biological applications of DDFT, which have a remarkable diversity. DDFT allows to understand biological systems across all scales. Ion channels \cite{Gillespie2008,GillespieXWM2005}, which can be found in cell membranes, are a small-scale biological system that can be modeled in DDFT. Moreover, DDFT has been used to model the membranes themselves \cite{StantonOCISLG2021}. Going to larger scales, we arrive at DDFT models of entire cells as used in applications to cancer growth \cite{ChauviereLC2012,AlSaediHAW2018}, microswimmers \cite{MenzelSHL2016,HoellLM2017,HoellLM2018,HoellLM2019,Loewen2021}, and bacteria \cite{WittmannSL2022}. In the SIR-DDFT model \cite{teVrugtBW2020,teVrugtBW2020b}, the considered \ZT{particles} are humans. It even does not have to stop there, since a (quantum-based) DFT has been applied to entire ecosystems \cite{TrappeC2021}. This brief list should make clear the particular advantage DDFT has in biology -- the same concept can be applied across all length scales, making DDFT an ideal tool for multiscale modeling.

\subsection{Chemistry}
When taking a look at the publications on DDFT from the past two years, it is notable that quite a number of them are in some way related to chemistry. Examples are the numerous applications of RDDFT \cite{teVrugtBW2020,teVrugtBW2020b,JeggleW2021,YiXJ2021,Loewen2021b,MonchoD2020,BleyDM2021,BleyHDM2022,AlstonPVB2022,LiuL2020,ZongWHWLH2022,Liu2020,ChenLL2022,TangYZQXZ2021} and the many works on electrochemistry \cite{QingZW2021,MaJLvR2022,QingJ2022,AslyamovSA2022,Frusawa2020,QingLZTQMXZ2020,Dejardin2022,MahdisoltaniG2021,AslyamovJ2022,Frusawa2022,Frusawa2022b,TomlinRKMG2021,ZhaoQLXZL2021,SuYZ2021,JiangZCHL2020,TaoLL2020,JeanmairetRS2022}. This is an interesting observation given that DDFT was developed as and is generally thought of as a theory for simple and colloidal fluids. 

Since this trend is a rather recent development, DDFT has a lot of unexplored potential for chemistry. Essentially, any system in which chemical reactions occur in combination with other interactions -- among the reactants or with other molecules in the environment -- could get an improved description from DDFT. This includes, in particular, many biochemical reactions which take place in crowded environments \cite{Minton2001}. Moreover, DDFT for ions can be used to improve the design of capacitors and batteries and in medical applications for studying ion channels. In the future, DDFT can therefore be expected to be relevant not only for basic research in statistical mechanics, but also for applications in biotechnology, nanotechnology, and chemical engineering.

\section{\label{conc}Conclusions}
In this article, we have summarized recent progress in the field of classical DDFT and outlined perspectives for the future. Interesting work remains to be done at the interface between DDFT and other closely related theories, namely PFC models and PFT. Moreover, DDFT has recently found quite a number of applications that are related to chemistry, which strongly suggests that this is a promising area for future work. Finally, DDFT is a powerful tool for the multiscale modeling of active and biological matter.

\acknowledgments{We thank Rudolf Haussmann, Hartmut L\"owen, Fabian Jan Schwarzendahl, and Ren{\'e} Wittmann for helpful discussions. M.t.V.\ thanks the Studienstiftung des deutschen Volkes for financial support. R.W.\ is funded by the Deutsche Forschungsgemeinschaft (DFG, German Research Foundation) -- 283183152.}

\bibliography{refs}
\end{document}